\documentstyle[aps,preprint]{revtex}
\begin{document}
\draft
\preprint{BARI - TH 238/96}
\date{June 24, 1996}
\title{
VACUUM STABILITY FOR DIRAC FERMIONS \\ IN  THREE DIMENSIONS }
\author{Paolo Cea}
\address{
Dipartimento di Fisica dell'Universit\`a di Bari, 
70126 Bari, Italy\\
{\rm and}\\
Istituto Nazionale di Fisica Nucleare, Sezione di Bari,
70126 Bari, Italy\\
(E-mail: cea@bari.infn.it)}
\maketitle
\begin{abstract}

\noindent
 I investigate three dimensional abelian and non-abelian gauge theories
 interacting with Dirac fermions. Using a variational method I evaluate
 the vacuum energy density in the one-loop approximation. It turns out
 that the states with a constant magnetic condensate lie below the
 perturbative ground state only in the case of three dimensional quantum
 electrodynamics with massive fermions.
\end{abstract}
\vspace{1cm}
\narrowtext
 More than ten years ago I showed that in three dimensional $QED$ with
 odd-Parity Dirac fermions the perturbative vacuum is unstable toward the
 formation of a uniform magnetic condensate~\cite{Cea85}. Moreover
 I also showed that that result holds in the presence of the topological mass
 term, and it is stable towards the Coulombic energy of the Dirac
 sea~\cite{Cea86}. 

 Recently there was a renewed interest in three dimensional 
 electrodynamic~\cite{Hosotani93,Cangemi95,Gusynin94,Gusynin96,Zeitlin95}.
 The interest has been triggered by the author of Ref.~\cite{Hosotani93}
 which reobtained the results of Refs.~\cite{Cea85,Cea86}. Moreover,
 it has been suggested that the vacuum of three dimensional $QED$ with
 massive Dirac fermions could be unstable for the presence
 of inhomogeneous magnetic field~\cite{Cangemi95}. It should be stressed
 that the results of Ref.~\cite{Cangemi95} apply to the case of massive 
 Dirac fermions with Parity invariant mass term. In the case of massive 
 Dirac fermions with Parity violating mass term the variational calculation
 of Ref.~\cite{Cea85} shows that:  
\begin{equation}
\label{eq:1}
 {\cal E}(B) \, = \,m \,\frac{|eB|}{4 \pi} \, + \, {\cal O}(B^2)
\end{equation}
 where ${\cal E}(B)$ is the vacuum energy density.

 Thus, for Dirac fermions with negative mass term $m=-|m|$ the states
 with constant magnetic field lie below the perturbative ground state.
 
 Similar considerations apply to the analysis performed in 
 Ref.~\cite{Gusynin94}. These authors claim that a constant magnetic
 field induces the dynamical flavor symmetry breaking. They used
 the reducible representation of the Dirac algebra and four-component
 spinors. It turns out that the approach of Ref.~\cite{Gusynin94}
 is relevant for the P- and T-invariant formulation of 
 electrodynamics~\cite{Footnote1}. The main result of Ref.~\cite{Gusynin94} 
 is that in the limit $m \, \rightarrow \, 0$ 
\begin{equation}
\label{eq:2}
 \langle0|\bar{\psi} \, \psi|0 \rangle \, = \,-\, \frac{|eB|}{2 \pi} \,\, .
\end{equation}
 However a closer look at Eq.(6) of Ref.~\cite{Gusynin94} shows that
 the correct result is~\cite{Footnote2}:
\begin{equation}
\label{eq:3}
 \langle0|\bar{\psi} \, \psi|0 \rangle \, = \,-\, \frac{|eB|}{2 \pi} \, 
 \frac{m}{|m|}  \, = \,- \, sign(m) \, \frac{|eB|}{2 \pi} \, \, .
\end{equation}
The mistake is due to the implicit assumption that $m \, \geq \, 0$.
 Note that Eq.~(\ref{eq:3}) is in accordance with the well known result
 that in the ground state of the massless theory there is a current of
 abnormal Parity~\cite{Alvarez83}:
\begin{equation}
\label{eq:4}
 \langle0|J^{\mu}(x)|0 \rangle \, =  \,
 \frac{m}{|m|}  \, \, \frac{e^2}{4 \pi}  \, \, ^*F^{\mu}(x) \, = \, 
 sign(m)  \, \, \frac{e^2}{4 \pi} \,  \, ^*F^{\mu}(x) 
\end{equation}
where $^*F^{\mu}(x) \,=\, \frac{1}{2}  \, \epsilon^{\mu\alpha\beta} 
 \, F_{\alpha\beta} (x) $.

From Equation~(\ref{eq:3}) it follows that the massless limit is ambiguous.
For instance, if one adopts the perfectly acceptable symmetric limit 
\begin{equation}
\label{eq:5}
 \frac{1}{2} \, \, ( \, \lim_{m \rightarrow 0^+} \, + \, 
 \lim_{m \rightarrow 0^-} \, ) \, \,
 \langle0|\bar{\psi} \, \psi|0 \rangle \, = \, \, 0 \,\, ,
\end{equation}
then one could conclude that there is no spontaneous symmetry breaking.

Instead, I will argue that there is dynamical symmetry breaking within the
Parity non invariant $QED$. To see this I consider the Parity non 
 invariant formulation of Ref.~\cite{Cea85} in an external constant magnetic
 field. Using the result of Ref.~\cite{Cea85}  it is easy to find
\begin{equation}
\label{eq:6}
 \langle 0 |\bar{\psi} \, \psi| 0 \rangle \, = \, \frac{|eB|}{2 \pi} \,
 \sum^{\infty}_{n=1} \frac{- \, m}{\sqrt{2|eB|n \,+\, m^2}} \, \, , \,  
  \, m \, > \, 0
\end{equation}
\begin{equation}
\label{eq:7}
 \langle0|\bar{\psi} \, \psi|0 \rangle \, = \, \frac{|eB|}{2 \pi} \,
 \sum^{\infty}_{n=0} \frac{- \, m}{\sqrt{2|eB|n \,+\, m^2}} \, \, , \,  
  m \, < \, 0 \, \, .
\end{equation}
 A straightforward calculation gives :
\begin{eqnarray}
\label{eq:8}
 \langle0|\bar{\psi} \, \psi|0 \rangle \, = && \,+ \, 
 \sqrt{\frac{|eB|}{2}} \, \frac{m}{4 \pi} \,\zeta \left( \frac{3}{2} \right)
 \, \, 
\nonumber \\
&& \, + \; {\cal O}\left( \frac{m^2}{|eB|} \right) \, \, , \,  m \, > \, 0
\end{eqnarray}
\begin{eqnarray}
\label{eq:9}
 \langle0|\bar{\psi} \, \psi|0 \rangle \, = && \, + \, \frac{|eB|}{2 \pi} \,
 + \,
 \sqrt{\frac{|eB|}{2}} \, \frac{m}{4 \pi} \,\zeta \left( \frac{3}{2} \right)
\nonumber \\ 
&& \,  + \; {\cal O} \left( \frac{m^2}{|eB|}\right) \, \, , \, m \, < \, 0 
 \, \, .
\end{eqnarray}
So that in the massless limit :
\begin{equation}
\label{eq:10}
  \lim_{m \rightarrow 0^+} \, \, \langle0|\bar{\psi} \, \psi 
 |0 \rangle \, =  \, 0 
 \end{equation}
\begin{equation}
\label{eq:11}
 \lim_{m \rightarrow 0^-} \, \, \langle0|\bar{\psi} \, \psi 
 |0 \rangle \, =  \,
 \, \frac{|eB|}{2 \pi}  \, \, .
\end{equation}
 Equation~(\ref{eq:11}) shows that the background magnetic field induces
 the fermionic condensate only in the case of negative Parity non invariant
 mass term. Moreover it turns out that in this last case the perturbative 
 ground state is unstable, for the state with an uniform magnetic field
 lies below the perturbative vacuum~\cite{Cea85}. Therefore in three
 dimensional $QED$ with massless fermions it is energetically
 favorable the formation of a uniform magnetic condensate and thereof the
 dynamical generation of a constant negative P-odd fermion mass.
 Note that the variational approach of Ref.\cite{Cea85} fixes only the 
 ratio $\frac{|eB|}{m^2 }$ as a function of $\frac{|m|}{e^2 }$. In order
 to determine separately the induced background field and the fermion
 mass one must generalize the variational approach of Ref.\cite{Cea85}
 by allowing  in the variational ground state wavefunctional the presence
 of both the background field and the fermion condensate.

 In the remainder of this paper I would like to show that this remarkable
 feature of three dimensional $QED$ does not extend to the case of
 fermions interacting with non Abelian gauge fields.

 Without loss in generality I consider the $SU(2)$ gauge theory with
 massive fermions. In the temporal gauge $A^a_0 =0$ the Hamiltonian is 
\begin{equation}
\label{eq:12}
 H = \, H_{Y-M} \, + \, H_D 
\end{equation}
where
\begin{equation}
\label{eq:13}
H_{Y-M} = \frac{1}{2} \int d^2x  \,\,
\left\{
\left( E^a_i(\vec{x}) \right)^2 +
\left( B^a(\vec{x}) \right)^2
\right\}   \; ,
\end{equation}
\begin{equation}
\label{eq:14}
 B^a(\vec{x}) \, = \, \frac{1}{2} \epsilon_{ij} F_{ij}^a (\vec{x}) \; ,
\end{equation}
\begin{equation}
\label{eq:15}
H_D  =  \int d^2x \, \psi^{\dagger}(\vec{x}) 
 \left\{
  \vec{\alpha} \cdot [ \, -i \vec{\nabla} \, - g \vec{A}(\vec{x}) \,]
 + \beta m 
 \right\} \psi(\vec{x})  \; .
\end{equation}
 In Equation~(\ref{eq:15}) $\vec{A} = \vec{A}^a \, \,\frac{\tau^a}{2}$ and 
 I use the two-dimensional realization of the Dirac algebra.
 In the fixed-time Schr\"odinger representation the chromoelectric
 field $E^a_i(\vec{x})$ acts as functional derivative on the physical
 states which are functionals obeying the Gauss'~law:
\begin{eqnarray}
\label{eq:16}
&& \left[ \partial_i \delta^{ab} + g \epsilon^{acb} A^c_i(\vec{x}) \right]
\frac{\delta}{\delta A^b_i(\vec{x}) } \, \, {\cal S}[A,\psi^\dagger,\psi] = 
\nonumber \\ 
&& i \, g \, \psi^{\dagger}(\vec{x}) \frac{\tau^a}{2} \psi(\vec{x}) 
 \, \, {\cal S}[A,\psi^\dagger,\psi] \; .
\end{eqnarray}
The effects of an external background field is incorporated by
writing
\begin{equation}
\label{eq:17}
A^a_i(\vec{x}) =  \bar{A}^a_i(\vec{x}) + \eta^a_i(\vec{x})
\end{equation}
with
\begin{equation}
\label{eq:18}
\bar{A}^a_i(\vec{x}) =  \delta^{a3} \delta_{i2} x_1 B \;, \; \,
\vec{x} = (x_1, x_2) \; .
\end{equation}
In the one-loop approximation the Gauss'~law reduces to:
\begin{equation}
\label{eq:19}
 D^{ab}_i(\vec{x}) \, 
\frac{\delta}{\delta \eta^b_i(\vec{x}) } \, {\cal S}[\eta,\psi^\dagger,\psi] = 
 i \,  g \, \psi^{\dagger}(\vec{x}) \frac{\tau^a}{2} \psi(\vec{x}) 
 \,  {\cal S}[\eta,\psi^\dagger,\psi]  
\end{equation}
 where $D^{ab}_i(\vec{x}) = \partial_i \delta^{ab}
 + g \epsilon^{acb} \bar{A}^c_i (\vec{x}) \,$ . The solution of 
 Eq.~(\ref{eq:19}) reads
\begin{equation}
\label{eq:20}
 {\cal S}[\eta,\psi^\dagger,\psi] \, = \, {\cal G}[\eta] \, \, 
 {\cal F}[\psi^\dagger,\psi] \; \exp \{ \Gamma[\eta,\psi^\dagger,\psi] \}
\end{equation}
where
\begin{equation}
\label{eq:21}
 D^{ab}_i(\vec{x}) \,
 \frac{\delta {\cal G}[\eta]}{\delta \eta^b_i(\vec{x})} \, = 0 \, \, ,
\end{equation}
i.e. the wavefunctional ${\cal G}[\eta]$ depends on the fluctuation fields
 which are transverse with respect to the background field covariant
 derivative. As concern the functional $\Gamma[\eta,\psi^\dagger,\psi]$,
 one can show~\cite{Cea85} that $\Gamma[\eta,\psi^\dagger,\psi]$ adds to
 the Hamiltonian the following Coulombic term:
\begin{equation}
\label{eq:22}
 H_C \, = \,
  \frac{g^2}{2}  \int d^2x \, d^2y  \,\, 
   \rho^{a}(\vec{x}) \,\, 
   ( D^{-2} )^{ab}(\vec{x},\vec{y}) \,\,
  \rho^{b}(\vec{y})
  \, ,
\end{equation}
where $(D^{2})^{ab} =  D^{ac}_k D^{cb}_k$ and 
 $\rho^{a}(\vec{x})=\psi^{\dagger}(\vec{x}) \, \frac{\tau^a}{2} \, 
 \psi(\vec{x})$. It is  obvious that in the one-loop approximation 
 the Coulombic energy can be neglected. Thus, the ground state energy 
 can be written as
\begin{equation}
\label{eq:23}
 E \; = \; E_{Y-M} \; + \; E_D \; 
\end{equation}
where
\begin{equation}
\label{eq:24}
  E_{Y-M} \; = \; \frac{ \langle{\cal G}|H_{Y-M} |{\cal G} \rangle }
 {\langle{\cal G} |{\cal G} \rangle }  \; ,
\end{equation}
\begin{equation}
\label{eq:25}
 E_D \; = \; \frac{ \langle{\cal F}|H_D |{\cal F} \rangle }
 {\langle{\cal F} |{\cal F} \rangle }  \; .
\end{equation}
Due to the presence of the Nielsen-Olesen unstable modes~\cite{Nielsen78}
 the calculation of $E_{Y-M}$ is quite involved even in the one-loop
 approximation. Nevertheless this calculation has been 
 performed~\cite{Cea93}. Here I merely quote the final result ($gB>0$):
\begin{equation}
\label{eq:26}
\frac{ E_{Y-M}(B)}{V} = + \frac{\sqrt{2} -1}{8 \pi^2}
\, \zeta\left( \frac{3}{2} \right)
\, (gB)^{3/2} \, + \,
{\cal O} ( g^2 (gB)^2 ) \, .
\end{equation}
 It is worthwhile to stress that the cancellation of the classical magnetic
 energy arises from the stabilization of the unstable modes. Moreover
 such dramatic effect has been checked by the Monte Carlo numerical
 simulations~\cite{Cea93}.

 In the one-loop approximation the Dirac energy $E_D$ is obtained by
 replacing in $H_D$ the dynamic gauge fields with the background field.
 As it is well known, in this case $E_D$ coincides with the energy of
 the Dirac sea. Then it is enough to solve the Dirac equation in presence
 of the background field Eq.~(\ref{eq:18}):
\begin{equation}
\label{eq:27}
  \{
  \vec{\alpha} \cdot [-i \vec{\nabla} \, - g \frac{\tau^a}{2} 
 \vec{\bar{A}^a}(\vec{x})] 
 + \beta m \, \} \psi(\vec{x})\, = \, E \, \psi(\vec{x}) \, .
\end{equation}
Introducing the $\tau^3$ eigenstates $\psi^{(\alpha)} \, , \alpha=1,2$, it is
 easy to see that Eq.~(\ref{eq:27}) reduces to two independent
 Abelian Dirac equations with charges $ \pm \frac{g}{2}$. It is, now,
 a straightforward exercise to solve Eq.~(\ref{eq:27}). In the case
 of negative mass term $m=-|m|$ I find: 
\begin{equation}
\label{eq:28}
  E^{(\alpha=1)}_{n,\pm} \, = \, \pm E_n \; ,
 \, n \geq 1 \, ; \;
  \, E^{(\alpha)}_{0}\, = \, - \, |m|
\end{equation}
\begin{equation}
\label{eq:29}
  E^{(\alpha=2)}_{n,\pm} \, = \, \pm E_{n+1} \; \, ,
  \, n \geq 0 \, \, ,
\end{equation}
where $E_{n}=\sqrt{gBn \,+\, m^2}$. Thus the vacuum energy is :
\begin{equation}
\label{eq:30}
 \frac{ E_{D}(B)}{V} \, = \,- \, \frac{gB}{4 \pi} \,
 \left\{  
  2 \, \sum^{\infty}_{n=1} \sqrt{gBn \,+\, m^2} \, \, + \, \, |m|  
 \right\}  \, \, .
\end{equation}
Subtracting the vacuum energy for $B=0$, and using the function~\cite{Cea85}
\begin{equation}
\label{eq:31}
 g(\lambda) \, = \,  \int^{\infty}_0 \frac{dx}{\sqrt{\pi} \sqrt{x}} \,\,
\frac{d}{dx} \, e^{-\frac{x}{\lambda}}
\left\{ \frac{1}{1 - e^{- 2 \, x}} - \frac{1}{2x} \right\} \; ,
\end{equation}
 I get:
\begin{equation}
\label{eq:32}
\frac{E_D (B)}{V} \, = \, \frac{(gB)^{\frac{3}{2}}}{2 \pi}
\, \left[ \frac{g(\frac{\lambda}{2})}{\sqrt{2}} 
  \, + \, \frac{1}{2 \sqrt{\lambda}}   \right] \; , \;
 \lambda \, = \, \frac{gB}{m^2}  
\end{equation}
where to perform the sum over $n$ I used the integral representation:
\begin{equation}
\label{eq:33}
\sqrt{a} = -\int^{\infty}_0 \frac{ds}{\sqrt{\pi} \sqrt{s}} \,\,
\frac{d}{ds} e^{-as}  \; \;, \; a \, > \, 0 \; .
\end{equation}
For completeness I report also the case of Parity invariant mass term:
\begin{equation}
\label{eq:34}
\frac{E_D (B)}{V} \, = \, \frac{(gB)^{\frac{3}{2}}}{2 \pi}
\, \left[ \frac{g(\frac{\lambda}{2})}{\sqrt{2}} 
  \, + \, \frac{1}{ \sqrt{\lambda}}   \right] \; .
\end{equation}
Note that for massless fermions both Eqs.~(\ref{eq:32}) and (\ref{eq:34}) 
 reduce to the known result~\cite{Trottier91}:
\begin{equation}
\label{eq:35}
\frac{E_D (B)}{V} \, = \, \frac{1}{8 \pi^2}
 \, \zeta\left( \frac{3}{2} \right) \, (gB)^{3/2} \; .
\end{equation}
Now, using~\cite{Cea85}
\begin{equation}
\label{eq:36}
 g( z ) \, \stackrel{z \rightarrow  0 }{\sim}  
 \, - \, \frac{1}{2 \sqrt{z}} \, + \frac{ \sqrt{z}}{12}  \;  ,  
\end{equation}
one obtains 
\begin{equation}
\label{eq:37}
 \frac{E_D (B)}{V} \;
 \stackrel{gB \rightarrow 0 }{\sim}  \, + \, \frac{1}{ 48 \pi} \, 
 \frac{ (gB)^2}{|m|} \; .
\end{equation}
This last equation should be contrasted with Eq.~(\ref{eq:1}).
 Adding the gauge field contribution I finally obtain:
\begin{equation}
\label{eq:38}
 {\cal E}(B) \equiv \frac{E (B)}{V} \, = 
  \frac{\sqrt{2} -1}{8 \pi^2}
 \, \zeta( \frac{3}{2})
 \, (gB)^{3/2} \, + \,
{\cal O} ( (gB)^2  )  \, .
\end{equation}
 Equation~(\ref{eq:38}) shows that the minimum of the vacuum energy density
 is attained for $B = 0 $. In other words, the perturbative ground state
 lies below the states with an uniform Abelian magnetic condensate.

 To conclude, using a variational approximation which is a refinement
 of the usual one-loop calculation I studied three dimensional
 gauge fields in interaction with Dirac fermions. The method developed
 in this paper allows to treat on the same footing Dirac fermions with both
 Parity invariant and non invariant mass term. Within the one-loop
 approximation the unique theory displaying a non trivial ground state
 turns out to be the three dimensional $U(1)$ gauge theory
 in interaction with  fermionic fields with negative P-odd mass term.
 Thereby, this result selects that model as the most promising model
 towards an effective description of planar systems in condensed matter
 physics.   
\end{document}